%% file: main.tex
\documentclass[10pt, conference, letterpaper]{IEEEtran}

\usepackage{graphicx} 
\usepackage[utf8]{inputenc}
\usepackage[T1]{fontenc}
\usepackage{microtype}
\usepackage{xargs}
\usepackage{lipsum}
\usepackage{xparse}
\usepackage{xifthen, xstring}
\usepackage{xspace}
\usepackage{marginnote}
\usepackage{etoolbox}
\usepackage{soul}
\usepackage{cite}
\usepackage{amsmath,amsfonts}
\usepackage{graphicx}
\usepackage{textcomp}
\usepackage{hhline}
\usepackage{multirow}
\usepackage{booktabs}
\usepackage{subcaption}
\usepackage{tabularray}
\usepackage{pifont}
\usepackage[dvipsnames]{xcolor}
\usepackage{float}
\usepackage{colortbl}
\usepackage{threeparttable}
\usepackage{tablefootnote}
\usepackage{adjustbox}
\usepackage{amssymb}
\usepackage{pifont} 
\usepackage{newtxtext}
\usepackage{xcolor}
\definecolor{Rosewater}{RGB}{220, 138, 120}
\definecolor{Flamingo}{RGB}{221, 120, 120}
\definecolor{Pink}{RGB}{234, 118, 203}
\definecolor{Mauve}{RGB}{136, 57, 239}
\definecolor{Red}{RGB}{210, 15, 57}
\definecolor{Maroon}{RGB}{230, 69, 83}
\definecolor{Peach}{RGB}{254, 100, 11}
\definecolor{Yellow}{RGB}{223, 142, 29}
\definecolor{Green}{RGB}{64, 160, 43}
\definecolor{Teal}{RGB}{23, 146, 153}
\definecolor{Sky}{RGB}{4, 165, 229}
\definecolor{Sapphire}{RGB}{32, 159, 181}
\definecolor{Blue}{RGB}{30, 102, 245}
\definecolor{Lavender}{RGB}{114, 135, 253}

\usepackage{tabularx}
\usepackage{multicol}

\usepackage{stfloats}
\usepackage{algorithm}
\usepackage{algpseudocode}
\usepackage{hyperref}

\newcommand{\xdmaname}{\mbox{\textit{Torrent}}\xspace}
\newcommand{\xdmasname}{\mbox{\textit{Torrent}s}\xspace}
\newcommand{\chainwrite}{\mbox{\textit{Chainwrite}}\xspace}

\title{\xdmaname: A Distributed DMA for Efficient and Flexible Point-to-Multipoint Data Movement}

\begin{document}


\author{
    \IEEEauthorblockN{Yunhao Deng$^*$, Fanchen Kong$^*$, Xiaoling Yi, Ryan Antonio, Marian Verhelst}
    \IEEEauthorblockA{
        \textit{MICAS-ESAT, KU Leuven} \\
        \{yunhao.deng, fanchen.kong, xiaoling.yi, ryan.antonio, marian.verhelst\}@esat.kuleuven.be
    }
}

\maketitle
\def\thefootnote{}\footnotetext{This project has been funded by the European Research Council (ERC)
under grant agreement No. 101088865, the Flanders AI Research Program,
and long term structural Methusalem funding by the Flemish Government.
 }\def\thefootnote{\arabic{footnote}}

\def\thefootnote{*}\footnotetext{Both authors contributed equally to this research.}\def\thefootnote{\arabic{footnote}}
\input{0-abstract}
\input{1-introduction}
\input{2-background}
\input{3-Architecture}
\input{4-evaluation}
\input{5-conclusion}

\newpage
\bibliographystyle{IEEEtran}
\bibliography{refs}

\end{document}

%% file: 0-abstract.tex
\begin{abstract}
The growing disparity between computational power and on-chip communication bandwidth is a critical bottleneck in modern Systems-on-Chip (SoCs), especially for data-parallel workloads like AI. Efficient point-to-multipoint (P2MP) data movement, such as multicast, is essential for high performance. However, native multicast support is lacking in standard interconnect protocols. Existing P2MP solutions, such as multicast-capable Network-on-Chip (NoC), impose
additional overhead to the network hardware and require modifications to the interconnect protocol, compromising scalability and compatibility.

This paper introduces \xdmaname, a novel distributed DMA architecture that enables efficient P2MP data transfers without modifying NoC hardware and interconnect protocol. \xdmaname conducts P2MP data transfers by forming logical chains over the NoC, where the data traverses through targeted destinations resembling a linked list. This \chainwrite mechanism preserves the P2P nature of every data transfer while enabling flexible data transfers to an unlimited number of destinations. To optimize the performance and energy consumption of \chainwrite, two scheduling algorithms are developed to determine the optimal chain order based on NoC topology. 

Our RTL and FPGA prototype evaluations using both synthetic
and real workloads demonstrate significant advantages in performance, flexibility, and scalability over network-layer multicast. Compared to the unicast baseline, \xdmaname achieves up to a $7.88\times$ speedup. ASIC synthesis on 16nm technology confirms the architecture's minimal footprint in area (1.2\%) and power (2.3\%). Thanks to the \chainwrite, \xdmaname delivers scalable P2MP data transfers with a small cycle overhead of 82CC and area overhead of 207 $\mu m^2$ per destination.


\end{abstract}

%% file: 1-introduction.tex
\section{Introduction}
\label{sec:introduction}
The widening imbalance between compute capabilities and on-chip communication requirement needs has emerged as a fundamental bottleneck in modern Systems-on-Chip (SoCs). While transistor density continues to increase following Moore's Law, middle- and far-reach interconnect densities have remained nearly constant \cite{bohr200930}, leaving on-chip bandwidth as a scarce and critical resource. Although contemporary SoCs integrate increasingly powerful accelerators~\cite{axelera_dimc_2024_isscc,nectar_ra_soc_2024_hc}, the performance of data-intensive workloads—such as Transformer-based large language models (LLMs)—is often constrained by memory bandwidth rather than processing elements~\cite{gholami2024ai}. Consequently, maximizing data reuse\cite{sze2020efficient} has become a central strategy for sustaining high accelerator performance. Data reuse can be achieved at both the accelerator level\cite{gemmini-dac} and the SoC level. One example of SoC-level data reuse is tiled matrix multiplication, where one operand is tiled and the other operand needs to be distributed to multiple accelerators, creating a point-to-multi-point (P2MP) data movement pattern.

\begin{figure}[tp]
    \centering
    \includegraphics[width=0.95\linewidth]{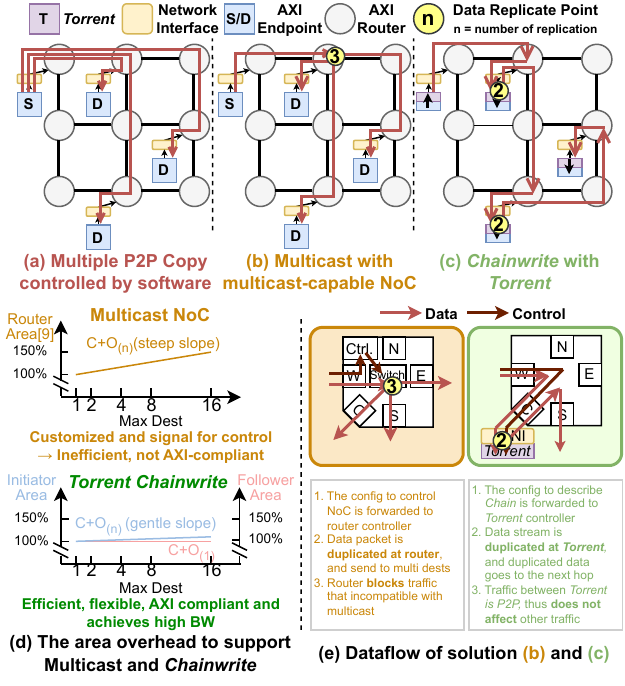}
    \caption{The working principle, area overhead, and the dataflow of three P2MP data copy mechanisms with a 2D Mesh NoC}
    \label{fig:intro}
\end{figure}
Software-based P2MP approaches~\cite{sw_multicast} are conducted by issuing multiple independent requests from the source (Fig.~\ref{fig:intro}(a)). However, these methods incur redundant memory accesses. To mitigate these inefficiencies, hardware support for multicast is being integrated into commercial~\cite{jang2021sparsity} and academic~\cite{esp_dma_2024} SoCs. By performing data replication within the interconnect (Fig.~\ref{fig:intro}(b)), hardware multicast avoids redundant memory traffic and improves data reuse. 

The Advanced eXtensible Interface (AXI) is a widely adopted on-chip communication protocol. Unfortunately, AXI lacks native support for P2MP communication~\cite{axi}. One approach to deploying multicast on top of AXI is to modify the Network-on-Chip (NoC) routers, equipping them with multicast capabilities as illustrated in Fig. \ref{fig:intro}(e). This allows a single input stream to be replicated to multiple output ports. While this approach is feasible, it requires protocol modifications and introduces extra control logic to prevent deadlocks\cite{eth_xbar_multicast}. More critically, as shown in Fig.~\ref{fig:intro}(d), the complexity of NoC routers and the link width grow with the maximum number of multicast destinations\cite{esp_dma_2024}. Thus, the scalability of network-layer multicast is often limited. 

In this paper, we propose \xdmaname \footnote{\href{https://github.com/KULeuven-MICAS/snax_cluster/tree/main/hw/chisel/src/main/scala/snax/xdma}{\textbf{\underline{\xdmaname Frontend}}} is open-sourced as one component in \href{https://github.com/KULeuven-MICAS/snax_cluster}{\textbf{{\underline{\textit{SNAX}}}}}, while \href{https://github.com/KULeuven-MICAS/xdma_axi_adapter}{\textbf{\underline{\xdmaname Backend}}} is open-sourced separately. }, an alternative architecture enabling efficient P2MP data transfer in AXI-compatible SoCs. Instead of altering the NoC infrastructure, we embed P2MP capabilities directly within the Direct Memory Access (DMA) endpoints (Fig. \ref{fig:intro}(c)(e)). When a P2MP request is sent to one \xdmaname (called the initiator \xdmaname), it collaborates with all other \xdmasname attached to destination memories (called follower \xdmasname). Together, they form a data chain\textemdash an architecture resembling a doubly linked list that allows data to flow from the first node to the last node. This multicast mechanism is called \chainwrite \footnote{In this paper, to ensure a clear terminology, the conventional P2MP data transfer with the help of a NoC router is referred to as \textit{Multicast} and our P2MP data transfer using \xdmaname is referred to as \chainwrite}. By shifting the data replication job from centralized NoC routers (at the network layer) to distributed \xdmasname (at the application layer), \xdmaname preserves full interoperability with AXI while allowing flexible P2MP capability. This approach supports a virtually unlimited maximal number of destinations ($N_{\text{dst,max}}$) without increasing router area or link width, while imposing ignorable hardware complexity (Fig. \ref{fig:intro}(d)).

Our main \textbf{contributions} are as follows:
\begin{itemize}
    \item We propose \xdmaname, a distributed DMA architecture that supports efficient and scalable P2MP data transfers at the application layer, a mechanism called \chainwrite.
    \item \chainwrite supports an unbounded number of destinations without modifying the AXI standard and NoC routers by organizing endpoints into a \textit{doubly linked list}.
    \item We develop scheduling algorithms that enable \chainwrite to be comparable with network-layer multicast in total number of hops for one P2MP data transfer task. 
    \item We validate \xdmaname using both synthetic and real workloads and compare against SoTA P2P and multicast solutions. Synthetic results show \xdmaname achieves P2MP efficiency on par with multicast solutions. The real workloads extracted from the DeepSeek-V3 self-attention layers are executed on a \xdmaname-enhanced SoC in an FPGA, showing up to $7.88\times$ speedup over its unicast baseline. ASIC synthesis in 16nm and power analysis results show \xdmaname incurs only 1.2\%/2.3\% of the system area/power. The hardware overhead for adding one maximal destination with \chainwrite is minimized to 207$\mu m^2$.
\end{itemize}

Compared to the SoTA DMAs and NoCs from industry and academia, as shown in Table \ref{tab:sota}, we observe that \xdmaname is distinctive from other P2MP mechanisms that use the NoC network layer to replicate data~\cite{esp_dma_2024}\cite{FlexNoC}\cite{eth_xbar_multicast}: by performing replication at the endpoints and using \chainwrite, \xdmaname achieves SoC-level data reuse without complicating the shared interconnect fabric. Furthermore, \xdmaname is highly flexible due to its distributed architecture, offering the ND affine access capability with low power and area overhead. 

\begin{table}[]
\resizebox{\columnwidth}{!}{%
\begin{tabular}{@{}lcccccc@{}}
\toprule
 &
  Arch. &
  \begin{tabular}[c]{@{}l@{}}Addr.\\ Gen    \end{tabular} &
  \begin{tabular}[c]{@{}l@{}}AXI  \\ Comp.  \end{tabular} &
  \begin{tabular}[c]{@{}l@{}}P2MP \\ Method \end{tabular} &
  \begin{tabular}[c]{@{}l@{}}Area \\ Scaling\end{tabular} &
  \begin{tabular}[c]{@{}l@{}}Open \\ Sourced\end{tabular} \\ \midrule
Xilinx DMA\cite{amd_dma}           & Mono. DMA & 1D & Yes & SW        & N/A          & No  \\
HyperDMA\cite{hyperdma_2024_icicm} & Dist. DMA & ND & No  & SW        & N/A          & No  \\
iDMA\cite{idma_paper}              & Mono. DMA & ND & Yes & SW        & N/A          & Yes \\
XDMA\cite{xdmav1}                  & Dist. DMA & ND & Yes & SW        & N/A          & Yes \\
\midrule
FlexNoC\cite{FlexNoC}              & NoC       & N/A& Yes & Multicast & N/A          & No  \\

ESP NoC\cite{esp_dma_2024}         & NoC       & N/A& No  & Multicast & O(N)         & Yes \\
Pulp XBar\cite{eth_xbar_multicast} & XBar      & N/A& Yes & Multicast & $\sim$O(1)   & Yes \\
\midrule
\textbf{\xdmaname}                 & \textbf{Dist. DMA} & \textbf{ND} & \textbf{Yes} & \textbf{\chainwrite} & \textbf{$\sim$O(1)} & \textbf{Yes} \\
\bottomrule
\end{tabular}%
}
\caption{\xdmaname comparison with SoTA DMAs and NoCs}
\label{tab:sota}
\end{table}

%% file: 2-background.tex
\section{Background}
\label{sec:background}
\subsection{Network-on-Chip Architecture}
Traditional hierarchical on-chip interconnects struggle to meet the performance requirements in complex multi-core SoCs, thus,  NoCs with flattened structures (like mesh, torus, etc.) have become appealing substitutes\cite{benini2002networks}. 

\begin{figure}[tp]
    \centering
    \includegraphics[width=\linewidth]{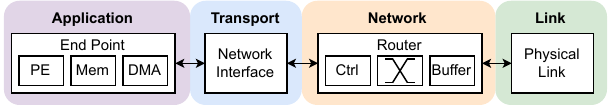}
    \caption{Layered on-chip network architecture}
    \label{fig:background_noc}
\end{figure}

NoCs can be segmented into four layers\cite{noc_layer}, referencing the OSI model (Fig.~\ref{fig:background_noc}): (1) The application layer includes hardware components that act as the requesters and responders for I/O requests; (2) The transport layer converts transactions from the application into routable packets and delivers them to the network layer; (3) The network layer forwards the packets to the port designated in the routing table; (4) The link layer is the physical P2P link between two routers to transfer data. 

\subsection{Multicast on NoC}
Similar to multicast in computer networks, the intuitive way to perform multicast on-chip is to rely on the NoC router. The multicast packet traverses the common four-stage\cite{samman2008multicast}\cite{zhao2022constellation} router pipelines where special handling occurs at each phase: In the Route Computation (RC) phase, the head flit determines multiple output ports based on the pre-configured multicast destination set. During Virtual Channel Allocation (VA), the multicast packet requests available virtual channels for each identified output port simultaneously, which may stall if some VCs are unavailable. In Switch Allocation (SA) and Switch Traversal (ST) phases, interconnects are altered to connect multiple output ports to a single input port, duplicating the packet multiple times. The same mechanism repeats until one packet is delivered to each 
destination. 

%% file: 3-architecture.tex
\section{\xdmaname Architecture}
\label{sec:architecture}
\begin{figure}[tp]
    \centering
    \includegraphics[width=0.95\linewidth]{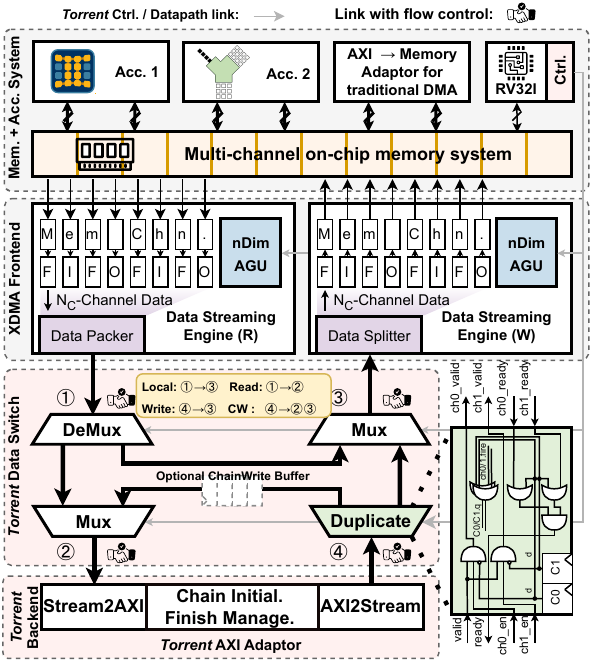}
    \caption{The architecture of \xdmaname}
    \label{fig:archi_main_architecture}
\end{figure}

\xdmaname proposes a novel decentralized architecture to conduct P2MP data transfers through \chainwrite. Unlike traditional DMAs~\cite{amd_dma}\cite{idma_paper} that perform internal loopback of two memory requests, \xdmaname adopts a distributed DMA architecture: multiple \xdmasname attached to the source and destination regions are involved in one data transfer task. 

The \xdmaname's microarchitecture is shown in Fig. \ref{fig:archi_main_architecture}: The \xdmaname Frontend is built on the open-source XDMA framework\cite{xdmav1} and its data streaming engine (DSE) \cite{yi2025datamaestro}, which can perform ND-affine memory accesses. This data
then enters the \xdmaname data switch, which duplicates and forwards the data to different ports. Finally, the \xdmaname Backend encapsulates data into AXI requests. 

\subsection{\xdmaname \chainwrite Orchestration}
\label{subsec:archi_cw_orchestration}
\begin{figure}[tp]
    \centering
    \includegraphics[width=0.95\linewidth]{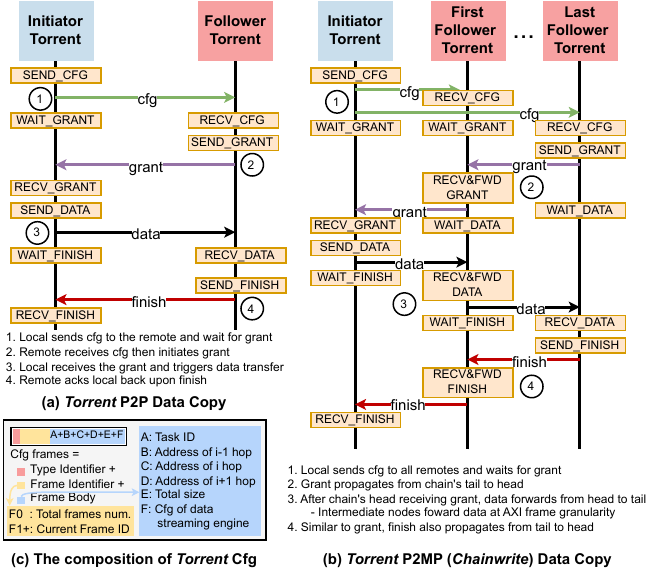}
    \caption{\xdmaname \chainwrite four-phase orchestration}
    \label{fig:archi_cw_orchestration}
\end{figure}

Since \xdmaname adopts a \textbf{distributed orchestration strategy} to accomplish a task, we design a dedicated control flow to avoid data corruption and deadlock. Specifically, we introduce a four-phase control flow, shown in Fig.~\ref{fig:archi_cw_orchestration} for \xdmaname P2P data copy and P2MP \chainwrite:

\textbf{(1) Configuration Dispatch}: When the initiator \xdmaname receives a P2P or P2MP task, it forwards the corresponding configuration settings (cfg, Fig.~\ref{fig:archi_cw_orchestration}(c)) to the involved remote \xdmasname. For P2P data copy, only one remote \xdmaname is involved; therefore, one cfg is sent out; for \chainwrite, cfgs are forwarded to all participating \xdmasname in parallel, with each cfg describing the address of the previous node and the next node. Hence, there is no theoretical limit on the length of the chain. These cfgs specify a doubly linked list on the SoC through which data can flow from
head to tail (for data) or tail to head (for control signals).

\textbf{(2) Grant Signaling Backward Propagation}: When the last \xdmaname  receives the cfg packet, it generates the Grant signal and sends it backward. Every intermediate \xdmaname only forwards the \textbf{Grant} to the previous node when it is ready for this new \chainwrite task.

\textbf{(3) Data Transfer}: Once the initiator \xdmaname receives the Grant signal, it begins sending data into the chain. Each intermediate \xdmaname stores and forwards every received data frame to the next hop as soon as it receives it from the previous hop, such that the data finally traverses through all \xdmasname. 

\textbf{(4) Finish Signaling Backward Propagation}: Similar to (2), a \textbf{Finish} signal is also generated by the last node and is propagated backward to the initiator \xdmaname to indicate that the \chainwrite on the chain has completed.

\subsection{The Cross-\xdmaname Configuration}
\xdmaname dispatches multi-field cfg packets (Fig.~\ref{fig:archi_cw_orchestration}(c)) to orchestrate with other \xdmasname: \textbf{Type Identifier} defines whether this frame is a read/write request; \textbf{Frame Identifier} describes the total number of frames (for the first frame) and current frame ID (for the remaining frames) of a cfg packet. The cfg is split into multiple \textbf{Frame Bodies} to support interconnect with variable lengths. Each frame body contains six fields: fields A to D describe the data chain; field E is for the \xdmaname Backend to generate AXI requests with corresponding size; and field F describes the access pattern for the DSE.

\subsection{The \xdmaname Datapath}
\label{subsec:archi_main_architecture}

The \textbf{\xdmaname Data Switch}, depicted in Fig.~\ref{fig:archi_main_architecture}, forwards and/or duplicates the data for different \xdmaname working modes: Local loopback, Read, Write, and \chainwrite. 

When the source and destination addresses are in the same memory, \xdmaname works in local loopback mode. In this case, \xdmaname is regarded as a dedicated data reshuffling accelerator, and the data flows from \ding{172} to \ding{174}. When the source and destination addresses are in different memories, then the \xdmaname attached to the source memory is in read mode (\ding{172} to \ding{173}) and the one attached to the destination memory is in write mode (\ding{175} to \ding{174}). In \chainwrite mode, the data switch duplicates the incoming data from \ding{175} into two copies; one copy is sent to the next hop (\ding{173}) and another copy is sent to the local DSE (\ding{174}). This behavior is reflected in the \textbf{RECV\&FWD DATA} State of Fig.~\ref{fig:archi_cw_orchestration}(b). The stream duplicator in the data switch enables the on-the-fly data duplication with no temporary storage of data, maximizing the energy efficiency of \chainwrite.

The \textbf{Backend} of \xdmaname serves as the bridge between the \xdmaname Frontend and the AXI interconnect, establishing lightweight “virtual tunnels” across \xdmasname on top of the AXI. 

\subsection{\xdmaname \chainwrite Sequence Algorithm}
\label{subsec:archi_scheduling_alg}

\begin{algorithm}
\caption{Chain Write Greedy Optimization Algorithm}
\label{alg:cw_greedy}
\small
\begin{algorithmic}[1]
\Require Destinations list $D$, NoC dimensions $noc_x$, $noc_y$
\Ensure Ordered traverse destination list $order$
\State $remaning_{dest} \gets D$ \Comment{Init remaining destinations}
\State $start \gets \min(remaning_{dest})$ \Comment{Start from dest closest to C0}
\State $order \gets [start]$, $remaning_{dest} \gets remaning_{dest} \setminus \{start\}$
\State $used_{path} \gets \text{XYpath}((0,0), start)$ \Comment{Initial path from C0}

\While{$remaing_{dest} \neq \emptyset$}
    \State $best \gets \text{null}$, $best_{hops} \gets noc_x + noc_y$
    \For{each $cand$ in $rem\_dest$}
        \State $path \gets \text{XYpath}(order[-1], cand)$
         \If{$\text{no\_overlap}(used_{path}, path)$ \& $|path| < best_{hops}$}
            \State $best \gets cand$, $best_{hops} \gets |path|$, $best_{path} \gets path$
        \EndIf
    \EndFor
    \If{$best = \text{null}$} \Comment{Fallback to shortest path}
        \State $best \gets argmin_{c \in remaning_{dest}} |\text{path}(order[-1], c)|$
        \State $best_{path} \gets \text{XYpath}(order[-1], best)$
    \EndIf
    \State $order.\text{append}(best)$
    \State $used_{path} \gets used_{path} \cup best_{path}$
    \State $remaning_{dest} \gets remaning_{dest} \setminus \{best\}$
\EndWhile
\State \Return $order$
\end{algorithmic}
\end{algorithm}

\chainwrite, in contrast to network-layer multicast, exposes destination traversal order explicitly. Our experiments demonstrate its significance for high performance. Thus, we design two complementary strategies to optimize this sequence:

\textbf{(1) Greedy heuristic (Alg.~\ref{alg:cw_greedy})} iteratively selects the next destination such that the routing path does not overlap with previously used links, while also minimizing path length. This approach balances efficiency with computational cost, making it well-suited for just-in-time optimization.

(2) The scheduling problem can be modeled as an open-path variant of the \textbf{Traveling Salesman Problem (TSP)}. Using Google OR-Tools~\cite{google_ortools}, we construct distance matrices based on XY-routing and solve for the globally optimal path. This approach guarantees a global optimum with higher computational overhead, making it a candidate for ahead-of-time optimization. 

%% file: 4-evaluation.tex
\section{Evaluation}
\label{sec:evaluation}
We evaluate \xdmaname in a multi-accelerator SoC. We first compare the P2MP efficiency of \xdmaname against the SoTA P2P DMA and the network-layer multicast solution (\S\ref{subsec:eval_cw}). We then demonstrate the effectiveness of the software \chainwrite sequence optimization to boost the P2MP performance (\S\ref{subsec:eval_cw_schedule}). Then, the configuration overhead is quantified in \S\ref{subsec:eval_overhead}, showing linear overhead scaling with the number of destinations. To validate the adaptability to real workloads, we prototype a multicore SoC on the AMD Versal\textsuperscript{\texttrademark} VPK180 FPGA and evaluate its performance in the self-attention layers of DeepSeek-V3 (\S\ref{subsec:eval_fpga}). Finally, we synthesize\footnote{Silicon synthesis targets TSMC 16FFC technology (600 MHz/0.8 V) using Synopsys Design Compiler\textsuperscript{\textregistered}. Power analysis is carried out on the synthesized netlist with gate-level switching activity via Synopsys PrimeTime\textsuperscript{\textregistered}.} a SoC with \xdmaname for area and power breakdowns (\S\ref{subsec:eval_asic}).

\subsection{System Evaluation Setup}
\label{subsec:eval_eval_setup}
In our system under evaluation, each compute cluster encompasses a 1MB, 32-bank, 64-bit-per-bank memory, two RV32I cores\cite{snitch_2020}, a GeMM accelerator with 1024 8-bit MACs, and a \xdmaname. 
The GeMM accelerator is optimized for LLM workloads and has two operating modes: (1) multiply two matrices of 16$\times$8 and 8$\times$8 (for the prefill stage); (2) multiply a vector of 1$\times$64 with a matrix of 64$\times$16 (for the decode stage). 
With this, we set up a 20-cluster SoC derived from Occamy\cite{paulin2024occamy}, interconnected by FlooNoC\cite{floonoc} with a $4\times5$ 2D mesh topology. The NoC uses XY-routing and the bandwidth is 64 bytes per cycle. 

Our primary comparison baseline is ESP\cite{esp_isscc}, a SoC platform with an in-house NoC that supports network-layer multicast\cite{esp_dma_2024}. For fair comparison, the ESP SoC is configured with the same 4$\times$5 2D mesh NoC, XY-routing, and 64 bytes/CC bandwidth. We additionally select iDMA\cite{idma_paper} for the software-based P2MP data copy condition.

\subsection{P2MP Copy Efficiency}
\label{subsec:eval_cw}
We evaluate the P2MP efficiency $\eta_{\text{P2MP}}$ under three approaches: (1) repeated P2P copies using iDMA, (2) Multicast in the NoC on ESP, and (3) \chainwrite with \xdmaname. Data sizes range from 1–128 KB and the number of destinations ($N_{\text{dst}}$) ranges from 2–16, yielding 192 test points.

The latencies are retrieved from hardware counters for all conditions. For iDMA, cycles equal the sum of all P2P transfers; \xdmaname measures the cycles from task dispatch to the DSE until the initiator \xdmaname receives the finish signal. Since the ESP platform only supports multicasting to accelerators, we implement dummy accelerators and integrate a hardware counter in the DMA  for the latency. 

Based on the measured latency\footnote{The iDMA and \xdmaname results are retrieved from RTL simulation. We conduct RTL simulation on ESP for sizes of 1-8KB and extrapolate the results for large tasks due to long simulation time.}, we define the P2MP efficiency $\eta_{\text{P2MP}}$ as follows:
\begin{equation}
\eta_{\text{P2MP}} = \frac{lat_{\text{P2P,theo}}}{lat_{\text{measured}}} = \frac{ N_{\text{dst}} \cdot \frac{Size_{data}}{BW_{\text{P2P,ideal}}} }{ lat_{\text{measured}} }
\end{equation}
where $lat_{\text{P2P,theo}} = N_{\text{dst}} \cdot \frac{Size_{data}}{BW_{\text{P2P,ideal}}}$ is the theoretical P2P copying latency for $N_{\text{dst}}$ destinations, and $BW_{\text{P2P,ideal}} = 64\text{B/CC}$ (system AXI bandwidth). By definition, iDMA will have $\eta_{\text{P2MP}}\le1$ since there is no data duplication and all data needs to be fetched from SRAM. Multicast or \chainwrite can reach $\eta_{\text{P2MP}}>1$ and the ideal condition is $\eta_{\text{P2MP}} = N_{\text{dst}}$, which means all destinations receive the data at the $BW_{\text{P2P,ideal}}$ with no delay. 

\begin{figure}[tp]
    \centering
    \includegraphics[width=0.85\linewidth]{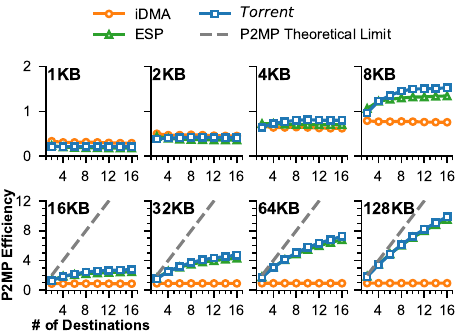}
    \caption{The $\eta_{\text{P2MP}}$ comparison for iDMA (Unicast), ESP (Multicast), and \xdmaname (\chainwrite)}
    \label{fig:eval_chainwrite}
\end{figure}

Fig. \ref{fig:eval_chainwrite} shows that $\eta_{\text{P2MP}}$ for \xdmaname and the ESP platform approach the ideal P2MP limit with increasing data sizes and $N_{\text{dst}}$. At small sizes (1KB–4KB), the control overhead dominates but is amortized with larger transfers. ESP outperforms \xdmaname for few-destination scenarios (2–4) due to lower link setup overhead, but its configuration complexity grows faster with $N_{\text{dst}}$ compared to \xdmaname. iDMA approaches $\eta_{\text{P2MP}}=1$ when transferring larger data (>8KB), indicating near-ideal P2P link utilization. This evaluation shows \xdmaname can achieve better performance than multicast in the NoC.

\subsection{Efficiency Impact of \chainwrite Sequence}
\label{subsec:eval_cw_schedule}
\begin{figure}[tp]
    \centering
    \includegraphics[width=0.85\linewidth]{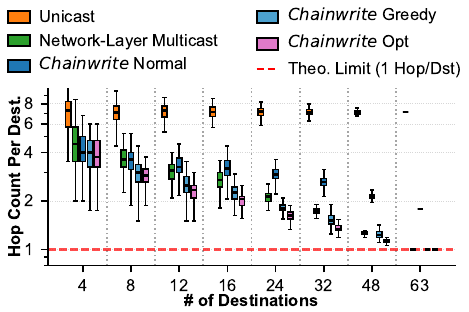}
    \caption{The comparison of average hops per destination}
    \label{fig:eval_cw_schedule}
\end{figure}

\chainwrite offers greater universality and flexibility than multicast, since it does not require modifications to AXI and can write data to different addresses with varying patterns. However, it is intuitively less efficient compared to multicast, in which data can be transferred along multiple paths in parallel instead of hop-by-hop. We compare \chainwrite against multicast by quantifying the average number of hops per destination, an implementation-agnostic metric that reflects energy consumption and latency, across all experiments. We set up an 8$\times$8-cluster mesh NoC and scan $N_{\text{dst}}$ from 4 to 63 (8 groups). Every group selects destinations randomly and repeats this 128 times (1024 test points in total). For example, one possible destination set for the $N_{\text{dst}}=4$ test initiated by $C_0$ is $\{C_3, C_7, C_{21}, C_{63}\}$. 

For unicast, the Manhattan distances between the source and all destinations are used to obtain the average number of hops. For multicast, one packet is routed following the standard XY-routing, and is divided when routes to different destinations do not overlap. For \chainwrite, we evaluate the two proposed scheduling algorithms introduced in \S \ref{subsec:archi_scheduling_alg} against a naive ordering following cluster IDs. Average hop counts are defined as the number of edges the data traverses divided by $N_{\text{dst}}$.

As shown in Figure~\ref{fig:eval_cw_schedule}, as $N_{\text{dst}}$ increases, unicast converges to the average Manhattan distance of the network while both multicast and \chainwrite converge to smaller values since packets exploit shared edges. Simple \chainwrite suffers from redundant paths and performs worse than multicast. However, \chainwrite with the greedy heuristic optimizer is comparable to multicast, while the TSP-based scheduling successfully helps \chainwrite surpass multicast at scale. For $N_{\text{dst}}=63$, both \chainwrite and multicast with optimizers converge to the theoretical limit of 1 hop per destination. 

\subsection{\chainwrite Configuration Overhead}
\label{subsec:eval_overhead}

\begin{figure}[tp]
    \centering
    \includegraphics[width=0.9\linewidth]{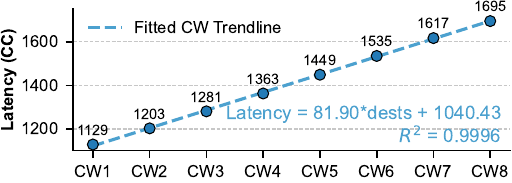}
    \caption{The configuration overhead to \chainwrite 64KB data to 1-8 destinations}
    \label{fig:eval_overhead}
\end{figure}

As discussed in \S\ref{subsec:archi_cw_orchestration}, \chainwrite introduces extra overhead due to the four-phase handshaking through the chain. To quantify the relationship between the overhead and the number of destinations, we measure latency for a 64KB data copy using \chainwrite with 1 to 8 destinations. 

Fig.~\ref{fig:eval_overhead} shows that \chainwrite setup incurs an overhead of 82 cycles per additional destination, following a linear trend. Although the latency value depends on the underlying NoC implementation, the scaling behavior remains consistent.

\subsection{DeepSeek-V3 Self-attention Layers Data Movement Evaluation on FPGA}
\label{subsec:eval_fpga}

\begin{figure}[tp]
    \centering
    \includegraphics[width=0.85\linewidth]{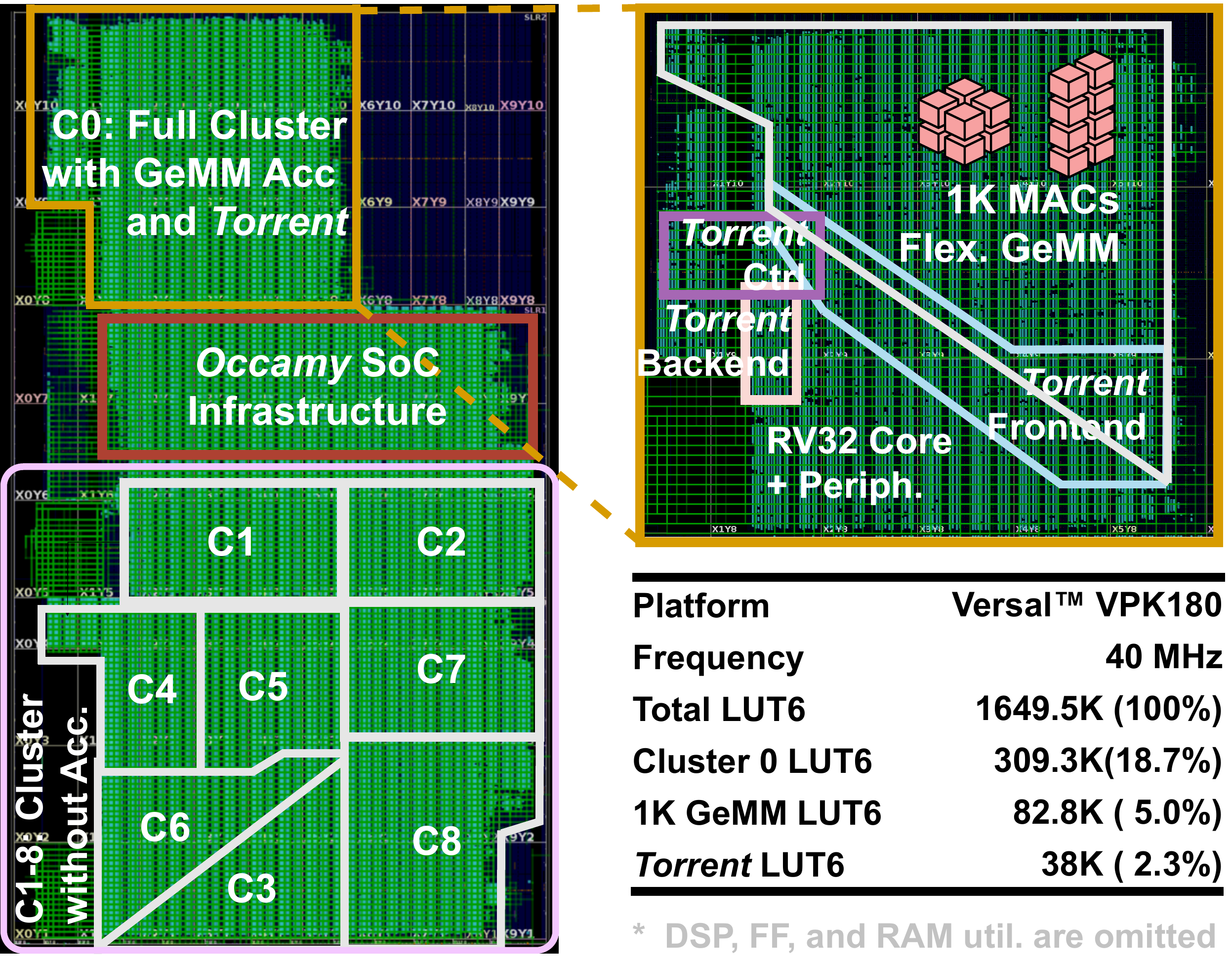}
    \caption{The FPGA implementation result of a 9-cluster SoC}
    \label{fig:eval_fpga}
\end{figure}

We implement the SoC with 3$\times$3 clusters on an AMD Versal\textsuperscript{\texttrademark} VPK180 FPGA. Among them, $C_0$ is a full cluster with the GeMM accelerator, \xdmaname, and XDMA; and the others are GeMM-less clusters due to FPGA resource constraints. Fig. \ref{fig:eval_fpga} details the annotated floorplan, frequency, and resource utilization of the FPGA. 

\begin{figure}[tp]
    \centering
    \begin{minipage}{\linewidth}
        \centering
        \includegraphics[width=0.9\linewidth]{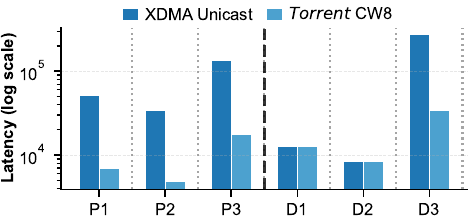}
        \vspace{-0.9em}
        \captionof{figure}{Latency of each configurations}
        \label{subfig:eval_deepseek_results}
        \vspace{0.5em}
    \end{minipage}
    \begin{minipage}{0.9\linewidth}
        \centering
        \resizebox{\linewidth}{!}{%
        \begin{tabular}{llll}
        \hline
        Experiement Setup  & Shape    & I/O Layout  & Multicast \\ \hline
        P1:QKT\_Single\_Head        & 2048$\times$192 & \begin{tabular}[c]{@{}l@{}}MNM16N8/\\ MNM8N8\end{tabular}   & TRUE      \\
        P2:SV\_Single\_Head          & 2048$\times$128 & \begin{tabular}[c]{@{}l@{}}MNM16N8/\\ MNM8N8\end{tabular}   & TRUE      \\
        P3:KV\_Matrix\_MLA\_Recovery & 2048$\times$512 & \begin{tabular}[c]{@{}l@{}}MNM16N8/\\ MNM16N8\end{tabular}  & TRUE      \\
        D1:QKT\_Single\_Head         & 4096$\times$192 & \begin{tabular}[c]{@{}l@{}}MNM16N8/\\ MNM64N16\end{tabular} & FALSE     \\
        D2:SV\_Single\_Head          & 4096$\times$128 & \begin{tabular}[c]{@{}l@{}}MNM16N8/\\ MNM64N16\end{tabular} & FALSE     \\
        D3:KV\_Matrix\_MLA\_Recovery & 4096$\times$512 & \begin{tabular}[c]{@{}l@{}}MNM16N8/\\ MNM16N8\end{tabular}  & TRUE      \\ \hline
        \end{tabular}%
        }
        \captionof{table}{Configuration Details}
        \label{subfig:eval_deepseek_config}
    \end{minipage}
    \caption{DeepSeekV3 Evaluation: Performance results (top) and configuration details (bottom)}
    \label{fig:eval_deepseek}
\end{figure}

We extract three workloads from the self-attention layers of Deepseek-V3\cite{liu2024deepseek} and the required data layouts from the architecture of the GeMM accelerator as shown in Table~\ref{subfig:eval_deepseek_config}: (1) When calculating $Q\cdot K^T$ at the prefill stage, the $K$ matrix (MNM16N8 layout) is the output of the previous matrix multiplication. Since the $Q$ matrix is large and will be tiled to multiple accelerators, the $K$ matrix needs to be copied to multiple accelerators; (2) The output of $Q\cdot K^T$ is then multiplied with the $V$ matrix. Thus, we need to transform its data layout and copy it to multiple destinations, similar to condition (1); (3) The recovery of the $KV$ matrix requires copying the KV-cache to all accelerators, but no layout transformation is required. The same workloads are also evaluated at the decode stage.

Fig.~\ref{subfig:eval_deepseek_results} shows that \xdmaname's \chainwrite P2MP data copying achieves up to 7.88$\times$ speedup over XDMA\cite{xdmav1} thanks to its highly-efficient ND affine access and \chainwrite capability. Overall, \xdmaname provides order-of-magnitude speedups in data copying operations of the self-attention layers, especially when P2MP and layout transformations are required. 

\subsection{ASIC Synthesis and Power Analysis}
\label{subsec:eval_asic}
\begin{figure}[tp]
    \centering
    \includegraphics[width=\linewidth]{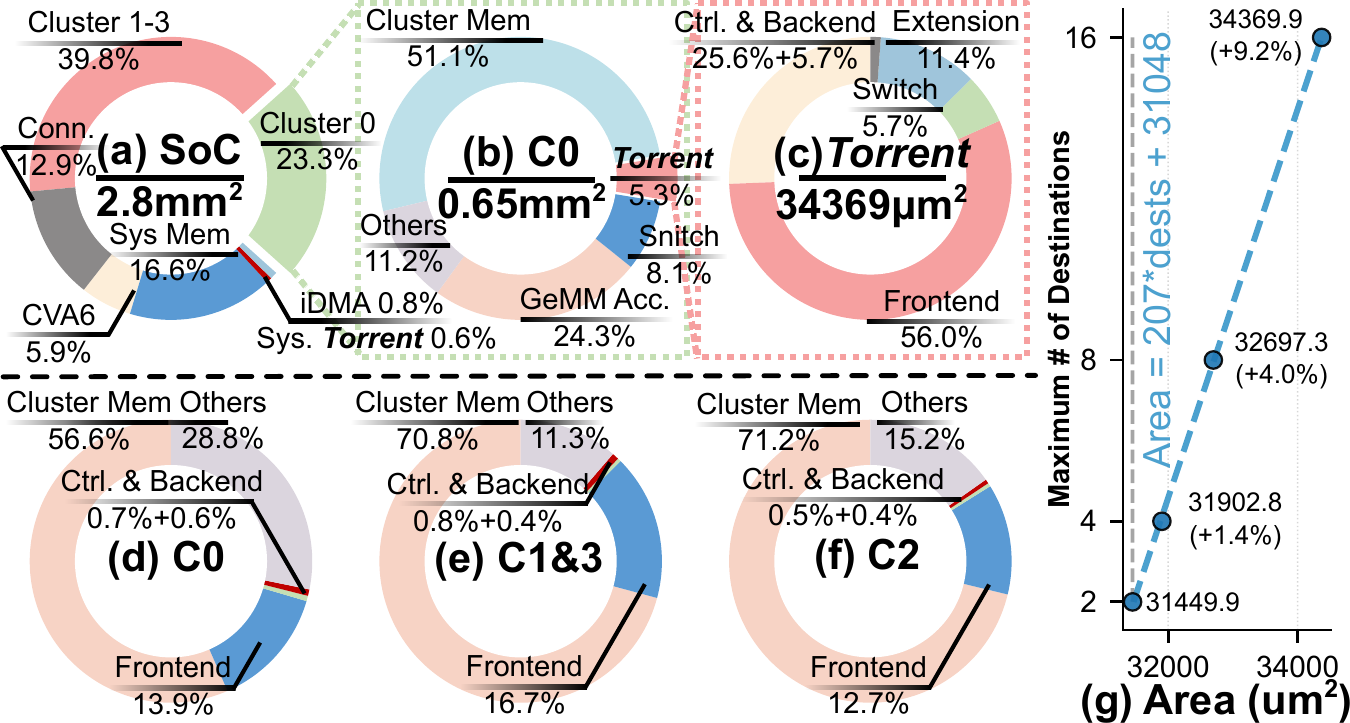}
    \caption{The area breakdown of (a) the 4-cluster SoC, (b) the accelerator cluster, and (c) the \xdmaname; the power breakdown of (d) the initiator cluster, (e) the follower cluster at the middle of the chain, and (f) the follower cluster at the tail of the chain; (g) the relationship between the area and maximum number of destinations for the initiator \xdmaname
    \label{fig:eval_area_power}}
\end{figure}

Finally, we synthesize the SoC in a 16 nm silicon technology, including 4 clusters with a 256KB scratchpad per cluster, and 512KB global memory. The first cluster is the fully-featured cluster with a GeMM accelerator, while the accelerator is removed from the remaining clusters. Five \xdmasname are instantiated (one attached to the global SRAM and one for each cluster) in the SoC. We conduct post-synthesis simulations with a 64KB, 3-destination \chainwrite from cluster 0. We also synthesize \xdmaname with different $N_{\text{dst,max}}$.

\subsubsection{Area Analysis}

Fig.~\ref{fig:eval_area_power}(a) shows the SoC consumes $2.8 mm^2$ of silicon area. Among them, the CVA6\cite{zaruba2019cost} core consumes 5.9\%, cluster 0 consumes 23.3\%, and the global SRAM consumes 16.6\% of the area. At the cluster scope (Fig.~\ref{fig:eval_area_power}(b)), \xdmaname consumes 5.3\% of the area, only equivalent to $\sim$1/5 of the size of the GeMM accelerator. \xdmaname has a comparable size to iDMA while providing efficient N-dimensional data copying and \chainwrite. At the SoC scope, the \xdmaname attached to global memory barely occupies 0.6\% of the area. Thanks to the \chainwrite architecture, the total \xdmaname area scales with a nearly constant hardware overhead of 0.65\% additional area per destination. (Fig. \ref{fig:eval_area_power}(g))

\subsubsection{Power Analysis}
The power of the initiator cluster is 175.7 mW (Fig.~\ref{fig:eval_area_power}(d)). Among the follower \xdmasname, the ones in the middle consume more power (Fig.~\ref{fig:eval_area_power}(e)) because they need to forward the data to the next hop. Our SoC reaches 4.68 pJ/B/hop energy efficiency. 

%% file: 5-conclusion.tex
\section{Conclusion}
\label{sec:conclusion}
We presented \xdmaname, a distributed DMA architecture that enables flexible application-layer multicast for large-scale SoCs. The FPGA implementation of \xdmaname-enhanced SoC demonstrates up to 7.88$\times$ speedup over unicast baseline. Compared to network-layer multicast, \xdmaname offers full AXI compatibility and greater flexibility. Software scheduling further improves efficiency and ensures full NoC bandwidth utilization. Hardware results confirm minimal overhead of 1.2\% in SoC area and 2.3\% in system power. \xdmaname delivers scalable P2MP data transfers with a small cycle overhead of 82CC and area overhead of 207 $\mu m^2$ per destination.

%% file: refs.bib
@article{idma_paper,
  title={A high-performance, energy-efficient modular DMA engine architecture},
  author={Benz, Thomas and Rogenmoser, Michael and Scheffler, Paul and Riedel, Samuel and Ottaviano, Alessandro and Kurth, Andreas and Hoefler, Torsten and Benini, Luca},
  journal={IEEE Transactions on Computers},
  volume={73},
  number={1},
  pages={263--277},
  year={2023},
  publisher={IEEE}
}

@article{esp_dma_2024,
  title={Towards Generalized On-Chip Communication for Programmable Accelerators in Heterogeneous Architectures},
  author={Zuckerman, Joseph and Wellman, John-David and Vanamali, Ajay and Shankar, Manish and Tombesi, Gabriele and Swaminathan, Karthik and Lee, Kevin and Kapur, Mohit and Philhower, Robert and Bose, Pradip and others},
  journal={arXiv preprint arXiv:2407.04182},
  year={2024}
}

@misc{xdmav1,
      title={XDMA: A Distributed, Extensible DMA Architecture for Layout-Flexible Data Movements in Heterogeneous Multi-Accelerator SoCs}, 
      author={Fanchen Kong and Yunhao Deng and Xiaoling Yi and Ryan Antonio and Marian Verhelst},
      year={2025},
      eprint={2508.08396},
      archivePrefix={arXiv},
      primaryClass={cs.AR},
      url={https://arxiv.org/abs/2508.08396}, 
}

@article{snitch_2020,
  title={Snitch: A tiny pseudo dual-issue processor for area and energy efficient execution of floating-point intensive workloads},
  author={Zaruba, Florian and Schuiki, Fabian and Hoefler, Torsten and Benini, Luca},
  journal={IEEE Transactions on Computers},
  volume={70},
  number={11},
  pages={1845--1860},
  year={2020},
  publisher={IEEE}
}

@inproceedings{paulin2024occamy,
  title={Occamy: A 432-core 28.1 DP-GFLOP/s/W 83\% FPU utilization dual-chiplet, dual-HBM2E RISC-V-based accelerator for stencil and sparse linear algebra computations with 8-to-64-bit floating-point support in 12nm FinFET},
  author={Paulin, Gianna and Scheffler, Paul and Benz, Thomas and Cavalcante, Matheus and Fischer, Tim and Eggimann, Manuel and Zhang, Yichao and Wistoff, Nils and Bertaccini, Luca and Colagrande, Luca and others},
  booktitle={2024 IEEE Symposium on VLSI Technology and Circuits (VLSI Technology and Circuits)},
  pages={1--2},
  year={2024},
  organization={IEEE}
}

@article{floonoc,
  title={FlooNoC: A 645-Gb/s/link 0.15-pJ/B/hop open-source NoC with wide physical links and end-to-end AXI4 parallel multistream support},
  author={Fischer, Tim and Rogenmoser, Michael and Benz, Thomas and G{\"u}rkaynak, Frank K and Benini, Luca},
  journal={IEEE Transactions on Very Large Scale Integration (VLSI) Systems},
  year={2025},
  publisher={IEEE}
}

@inproceedings{esp_isscc,
  title={14.5 A 12nm Linux-SMP-Capable RISC-V SoC with 14 accelerator types, distributed hardware power management and flexible noc-based data orchestration},
  author={Dos Santos, Maico Cassel and Jia, Tianyu and Zuckerman, Joseph and Cochet, Martin and Giri, Davide and Loscalzo, Erik Jens and Swaminathan, Karthik and Tambe, Thierry and Zhang, Jeff Jun and Buyuktosunoglu, Alper and others},
  booktitle={2024 IEEE International Solid-State Circuits Conference (ISSCC)},
  volume={67},
  pages={262--264},
  year={2024},
  organization={IEEE}
}

@software{google_ortools,
  title = {CP-SAT},
  version = { v9.12 },
  author = {Laurent Perron and Frédéric Didier},
  organization = {Google},
  url = {https://developers.google.com/optimization/cp/cp\_solver/},
  date = { 2025-02-17 }
}

@article{liu2024deepseek,
  title={Deepseek-v3 technical report},
  author={Liu, Aixin and Feng, Bei and Xue, Bing and Wang, Bingxuan and Wu, Bochao and Lu, Chengda and Zhao, Chenggang and Deng, Chengqi and Zhang, Chenyu and Ruan, Chong and others},
  journal={arXiv preprint arXiv:2412.19437},
  year={2024}
}

@inproceedings{hyperdma_2024_icicm,
  title={HyperDMA: Enhancing High-Performance Computing and AI Workflows with Advanced Data Transfer Capabilities},
  author={Peng, Minghao and Chen, Haiyan and Zhang, Yang and Liu, Sheng},
  booktitle={2024 9th International Conference on Integrated Circuits and Microsystems (ICICM)},
  pages={636--644},
  year={2024},
  organization={IEEE}
}

@article{bohr200930,
  title={A 30 year retrospective on Dennard's MOSFET scaling paper},
  author={Bohr, Mark},
  journal={IEEE Solid-State Circuits Society Newsletter},
  volume={12},
  number={1},
  pages={11--13},
  year={2009},
  publisher={IEEE}
}

@article{gholami2024ai,
  title={Ai and memory wall},
  author={Gholami, Amir and Yao, Zhewei and Kim, Sehoon and Hooper, Coleman and Mahoney, Michael W and Keutzer, Kurt},
  journal={IEEE Micro},
  volume={44},
  number={3},
  pages={33--39},
  year={2024},
  publisher={IEEE}
}

@inproceedings{jang2021sparsity,
  title={Sparsity-aware and re-configurable NPU architecture for Samsung flagship mobile SoC},
  author={Jang, Jun-Woo and Lee, Sehwan and Kim, Dongyoung and Park, Hyunsun and Ardestani, Ali Shafiee and Choi, Yeongjae and Kim, Channoh and Kim, Yoojin and Yu, Hyeongseok and Abdel-Aziz, Hamzah and others},
  booktitle={2021 ACM/IEEE 48th Annual International Symposium on Computer Architecture (ISCA)},
  pages={15--28},
  year={2021},
  organization={IEEE}
}

@inproceedings{axelera_dimc_2024_isscc,
  title={11.3 Metis AIPU: A 12nm 15TOPS/W 209.6 TOPS SoC for Cost-and Energy-Efficient Inference at the Edge},
  author={Hager, Pascal Alexander and Moons, Bert and Cosemans, Stefan and Papistas, Ioannis A and Rooseleer, Bram and Van Loon, Jeroen and Uytterhoeven, Roel and Zaruba, Florian and Koumousi, Spyridoula and Stanisavljevic, Milos and others},
  booktitle={2024 IEEE International Solid-State Circuits Conference (ISSCC)},
  volume={67},
  pages={212--214},
  year={2024},
  organization={IEEE}
}

@article{sw_multicast,
  title={Optimal software multicast in wormhole-routed multistage networks},
  author={Xu, Hong and Gui, Yadong and Ni, Lionel M},
  journal={IEEE Transactions on Parallel and Distributed Systems},
  volume={8},
  number={6},
  pages={597--607},
  year={1997},
  publisher={IEEE}
}

@misc{axi,
	author = {ARM},
	title = {AMBA AXI and ACE Protocol Specification},
	howpublished = {https://developer.arm.com/documentation/ihi0022/hc/?lang=en},
	year = {2021}
}

@INPROCEEDINGS{gemmini-dac,
  author={Genc, Hasan and Kim, Seah and Amid, Alon and Haj-Ali, Ameer and Iyer, Vighnesh and Prakash, Pranav and Zhao, Jerry and Grubb, Daniel and Liew, Harrison and Mao, Howard and Ou, Albert and Schmidt, Colin and Steffl, Samuel and Wright, John and Stoica, Ion and Ragan-Kelley, Jonathan and Asanovic, Krste and Nikolic, Borivoje and Shao, Yakun Sophia},
  booktitle={Proceedings of the 58th Annual Design Automation Conference (DAC)}, 
  title={Gemmini: Enabling Systematic Deep-Learning Architecture Evaluation via Full-Stack Integration}, 
  year={2021},
  volume={},
  number={},
  pages={}
}

@article{noc_layer,
  title={A survey of research and practices of network-on-chip},
  author={Bjerregaard, Tobias and Mahadevan, Shankar},
  journal={ACM Computing Surveys (CSUR)},
  volume={38},
  number={1},
  pages={1--es},
  year={2006},
  publisher={ACM New York, NY, USA}
}

@book{sze2020efficient,
  title={Efficient processing of deep neural networks},
  author={Sze, Vivienne and Chen, Yu-Hsin and Yang, Tien-Ju and Emer, Joel S},
  year={2020},
  publisher={Springer}
}

@article{eth_xbar_multicast,
  title={A Multicast-Capable AXI Crossbar for Many-core Machine Learning Accelerators},
  author={Colagrande, Luca and Benini, Luca},
  journal={arXiv preprint arXiv:2502.19215},
  year={2025}
}

@article{benini2002networks,
  title={Networks on chips: A new SoC paradigm},
  author={Benini, Luca and De Micheli, Giovanni},
  journal={computer},
  volume={35},
  number={1},
  pages={70--78},
  year={2002},
  publisher={IEEE}
}

@article{yi2025datamaestro,
  title={DataMaestro: A Versatile and Efficient Data Streaming Engine Bringing Decoupled Memory Access To Dataflow Accelerators},
  author={Yi, Xiaoling and Deng, Yunhao and Antonio, Ryan and Kong, Fanchen and Paim, Guilherme and Verhelst, Marian},
  journal={arXiv preprint arXiv:2504.14091},
  year={2025}
}

@inproceedings{nectar_ra_soc_2024_hc,
  title={NeCTAr and RASoC: Tale of Two Class SoCs for Language Model Interference and Robotics in Intel 16},
  author={Schmulbach, Viansa and Kim, Jason and Gao, Ethan and Jha, Nikhil and Wu, Ethan and Yu, Oliver and Oliveau, Ben and Kong, Xiangwei and Roberts, Brendan and McMahon, Connor and others},
  booktitle={2024 IEEE Hot Chips 36 Symposium (HCS)},
  pages={1--1},
  year={2024},
  organization={IEEE Computer Society}
}

@misc{amd_dma,
	author = {AMD Xilinx},
	title = {AXI DMA LogiCORE IP Product Guide},
	year = {2022}
}

@misc{FlexNoC,
	author = {},
	title = {{F}lex{N}o{C} {I}nterconnect {I}{P} - {A}rteris --- arteris.com},
	howpublished = {\url{https://www.arteris.com/products/non-coherent-interconnect-ip/flexnoc/}},
	year = {},
	note = {[Accessed 29-08-2025]},
}

@article{zaruba2019cost,
   author={F. {Zaruba} and L. {Benini}},
   journal={IEEE Transactions on Very Large Scale Integration (VLSI) Systems},
   title={The Cost of Application-Class Processing: Energy and Performance Analysis of a Linux-Ready 1.7-GHz 64-Bit RISC-V Core in 22-nm FDSOI Technology},
   year={2019},
   volume={27},
   number={11},
   pages={2629-2640},
   doi={10.1109/TVLSI.2019.2926114},
   ISSN={1557-9999},
   month={Nov},
}

@inproceedings{samman2008multicast,
  title={Multicast parallel pipeline router architecture for network-on-chip},
  author={Samman, Faizal A and Hollstein, Thomas and Glesner, Manfred},
  booktitle={Proceedings of the conference on Design, automation and test in Europe},
  pages={1396--1401},
  year={2008}
}

@inproceedings{zhao2022constellation,
  title={Constellation: An open-source SoC-capable NoC generator},
  author={Zhao, Jerry and Agrawal, Animesh and Nikolic, Borivoje and Asanovi{\'c}, Krste},
  booktitle={2022 15th IEEE/ACM International Workshop on Network on Chip Architectures (NoCArc)},
  pages={1--7},
  year={2022},
  organization={IEEE}
}
